\documentclass[12pt]{article}
\usepackage{epsf}
\usepackage{amsmath}
\pdfoutput=1
\usepackage{graphicx}
\usepackage{subfigure}
\usepackage{cite}

\renewcommand{\thefootnote}{\#\arabic{footnote}}
\begin{document}

\newcommand{\gtrsim}{ \mathop{}_{\textstyle \sim}^{\textstyle >} }
\newcommand{\lesssim}{ \mathop{}_{\textstyle \sim}^{\textstyle <} }

\newcommand{\rem}[1]{{\bf #1}}

\renewcommand{\thefootnote}{\fnsymbol{footnote}}
\setcounter{footnote}{0}
\begin{titlepage}

\def\thefootnote{\fnsymbol{footnote}}

\begin{center}
\hfill November 2016\\
\hfill REVISED
\vskip .5in
\bigskip
\bigskip
{\Large \bf A new direction for dark matter research: intermediate-mass compact halo objects}

\vskip .45in

{\bf George F. Chapline$^{(a)}$\footnote{email:george.chapline@gmail.com} and Paul H. Frampton$^{(b)}$\footnote{email:paul.h.frampton@gmail.com  homepage:paulframpton.org}}

\bigskip

{\em $^{(a)}$  Lawrence Livermore National Laboratory, P. O. Box 808, Livermore, CA, USA}

{\em $^{(b)}$ 15 Summerheights, 29 Water Eaton Road, Oxford OX2 7PG, UK}

\end{center}

\vskip .4in
\begin{abstract}
\noindent
The failure to find evidence for elementary particles that could serve as the constituents 
of dark matter brings to mind suggestions that dark matter might consist of massive 
compact objects (MACHOs). In particular, it has recently been argued that 
MACHOs with masses $> 15M_\odot$ may have been prolifically produced 
at the onset of the big bang. Although a variety of astrophysical signatures 
for primordial MACHOs with masses in this range have been discussed in the 
literature, we favor a strategy that uses the potential for magnification of stars 
outside our galaxy due to gravitational micro-lensing of these stars by MACHOs 
in the halo of our galaxy. We point out that the effect of the motion of the Earth 
on the shape of the micro-lensing brightening curves provides a promising 
approach to testing over the course of next several years the hypothesis 
that dark matter consists of massive compact objects.
\end{abstract}

\end{titlepage}

\renewcommand{\thepage}{\arabic{page}}
\setcounter{page}{1}
\renewcommand{\thefootnote}{\#\arabic{footnote}}

\newpage

\section{Introduction}

\noindent 
Astronomical observations have led to a consensus that the energy make-up of 
the visible universe is approximately $70\%$ dark energy (DE), $25\%$ dark matter (DM) 
and $5\%$ normal matter \cite{planck}. While DE apparently represents a vacuum energy, the 
physical nature of DM remains mysterious \cite{sanders}. Up to the present time the most 
popular theories of DM have been that it consists of either weakly interacting 
elementary particles with a mass between 1 GeV and 1 TeV (WIMPs) or very weakly 
interacting invisible axions \cite{bertone}. 

\bigskip

\noindent
The hypothesis that DM consists of WIMPs is plausible in the sense that the existence 
of WIMP-like particles is expected in any fundamental theory of elementary particles 
that possesses a supersymmetry between bosons and fermions. Supersymmetry gives rise 
to an R-parity which is needed to prevent unobserved proton decay and gave rise 
to the expectation that a light R=-1 particle, the neutralino, is the WIMP responsible 
for DM. Unfortunately, all attempts to identify DM WIMPs have so far failed.  Most notably 
at the CERN LHC, where supersymmetry was widely expected to make an 
appearance if dark matter consists of WIMPs, there is no sign of any WIMP; 
the best limits on WIMP detection are from liquid xenon detectors including
LUX \cite{LUX} and PANDA-X \cite{PANDAX}. Apparently supersymmetry is broken at an 
energy scale not accessible to the LHC \cite{peshkin}. Another type of elementary particle 
that has been much discussed as a candidate for DM is the invisible axion, 
with a mass $10^{-6} -10^{-3}$ eV. Because it appeared that it might be possible 
to detect a dark matter density of invisible axions using their signature conversion to 
2 photons in a strong magnetic field, clever experiments were initiated to search for 
DM axions. However, to date no evidence for invisible axions has appeared,
although it should be emphasized that only a modest fraction of the allowed mass
range has been scanned. 
Taken together with basic theoretical questions about the axion idea, a reasonable 
conclusion at this time is that alternatives to elementary particle models for DM 
should be seriously considered \cite{frampton}. In this paper we consider the alternative 
idea that DM consists of massive primordial compact objects in the form of 
either black holes \cite{hawking}, gravastars \cite{mazur}, or 
dark energy stars \cite{chapline}.

\bigskip

\noindent
The first dedicated searches for dark matter massive compact objects within our galaxy 
(MACHOs), carried out in the 1990's \cite{alcock,tisserand}, used the gravitational microlensing 
technique, so-called because of the very small angular sizes of the lenses as seen 
from the earth, first suggested by Paczynski \cite{paczynski} and Griest \cite{griest}. Although a 
number of microlensing events were recorded, the observed number of such events 
could not account for the total density dark matter known to exist in the Milky Way. 
The original MACHO search looked for events with durations ranging between 
about two hours and two hundred days, corresponding to MACHO masses between 
approximately $10^{-6} M_\odot$ and $15M_\odot$. However, to our knowledge, 
there is no fundamental reason why the technique of looking for gravitational 
microlensing used in the original \cite{alcock,tisserand} and subsequent MACHO 
searches \cite{calchi,wyrzykoeski} 
cannot be extended to higher mass MACHOs.

\bigskip

\noindent
MACHOs were originally conceived of as the end products of stellar evolution, 
i.e. white dwarfs, neutron stars, black holes, -- or possibly brown dwarfs -- and as a result of 
this prejudice it was not expected that MACHO masses would exceed $\sim 15 M_\odot$, 
corresponding to the mass of a black hole formed by the collapse of the core of a 
very massive star whose nuclear fuel has been exhausted. On the other hand it was 
suggested many years ago by one of us \cite{chapline1} that DM might consist of primordial 
black holes (PBHs). Although it was initially thought that these PBHs would typically 
have masses much smaller than the mass of the Sun \cite{hawking}, theoretical arguments 
have been advanced \cite{frampton2,frampton3,frampton4}
that PBHs with masses $>15M_\odot$ might have been 
prolifically produced at the onset of the big bang. Indeed there are cosmological models 
where primordial production of a DM density of either PBHs \cite{frampton5,carr}
or dark energy stars \cite{chapline2} 
with masses $>15M_\odot$ is favored. We will continue to use the acronym MACHO for 
these objects because for masses $< 10^5M_\odot$ such objects can contribute to the 
dark matter halo of our galaxy. In the following we will use the acronym IM MACHO to 
denote MACHOs with masses in the range 

\begin{equation}
15 M_\odot <  M_{MACHO}  <  10^5 M_\odot 
\end{equation}

\noindent 
i.e. intermediate between the masses of black holes formed as a result of stellar 
evolution and the supermassive compact objects found at the centers of galaxies.  
Although the 1990s search concluded that number of galactic MACHOs with masses 
$<15M_\odot$ was insufficient to explain the density of DM in our galaxy, we believe 
that the time is ripe to reconsider the idea that dark matter consists of MACHOs, 
particularly in the mass range (1). It might be noted that the recent LIGO detection 
of gravitational radiation from coalescing $30 M_\odot$ black holes is entirely 
consistent with the hypothesis that DM is made up of objects in this mass 
range \cite{bird}. In section 2 we will consider the question whether there are other 
astrophysical observations that bear on whether DM consists of IM MACHOs. 
Our general conclusion is that the question as to whether DM consists of IM MACHOs 
is best settled by using gravitational micro-lensing to limit the density of IM MACHOs 
in our galaxy. 

\bigskip

\noindent
Our main objective in this paper is to point out how the microlensing search techniques 
used in previous MACHO searches might be extended to definitively test over the 
next several years the hypothesis that the DM mass density is dominated by MACHOs 
with masses in the range in Eq.(1). Our strategy for directly detecting these MACHOs, 
discussed in more detail in Section 3 is to search for a transient brightening of stars 
beyond our galaxy that is consistent with gravitational micro-lensing, and then follow 
these candidate microlensing events over a period of a year or more to observe the 
parallax effect of the earth's motion on the microlensing. In section 3 we comment on 
how well this type of observing program maps onto the capabilities of astronomical resources 
that are available either currently or in the immediate future. 

\section{Can DM consist of IM MACHOs?}

\bigskip

\noindent
A well known theoretical argument that DM cannot consist of IM MACHOs with masses 
in the range Eq.(1) hinges on the consequences of accretion of interstellar gas onto a DM 
density of primordial MACHOs just after the time when electrons and nuclei in the interstellar 
gas recombine to form atoms$(z_{rec} \approx 10^3)$. The X-rays emitted by the accretion of 
this interstellar gas onto a DM population would heat the interstellar gas, which in turn would 
cause via the inverse Compton effect a distortion in the CMB both with regard to its spectrum 
and isotropy. A widely cited attempt to calculate this effect [23] employed the Bondi-Hoyle 
model for spherical accretion onto black holes, and carries through the computation all the 
way up to a point of comparison with limits on CMB spectral distortions derived from the 
COBE satellite observations. In particular it is claimed in \cite{ricotti} that the observed limits 
on the CMB distortion constrain the density of massive PBHs to only a tiny fraction $< 10^{-4}$ 
of the DM density. However the Bondi-Hoyle model for accretion ignores angular 
momentum conservation. Although this may be a good approximation before recombination 
when the angular momentum of accreting matter is rapidly dissipated by Compton viscosity \cite{loeb}, 
it becomes a questionable approximation after $z_{rec} \approx 10^3$ when the DM mass-energy 
density is already much larger than that of the CMB. Indeed it is well known \cite{kuo} that in the 
observable universe the Bondi-Hoyle model fails badly as a model for accretion onto the 
super-massive compact objects at the centers of galaxies - which in fact could themselves be 
primordial in origin. It was also noted in \cite{ricotti} that the limits on the contribution of massive 
PBHs to DM is sensitive to the fraction of time that the accretion onto the black hole is 
interrupted due to the Eddington limit for accretion. However if the IM MACHO is a dark 
energy star rather than a black hole, the intermittency of the accretion onto the primordial MACHOs 
would be significantly increased due to a dramatic lowering of the accretion rate where the 
Eddington limit is reached.  

\bigskip

\noindent
It has also been suggested that the existence of binary stars with wide separations might 
exclude \cite{Yoo} the existence of a DM density of IM MACHOs; however, the argument is indirect, 
and at the present time it is uncertain whether the observed abundance of wide binaries 
is inconsistent with a DM density of IM MACHOs \cite{quinn}. One argument that 
can be interpreted to be supportive of the 
hypothesis that DM does consist of IM MACHOs is that numerical simulations
\cite{frenk} of the 
spontaneous formation of inhomogeneous DM cosmic structures after $Z \sim 3000$, 
which typically use as a model for DM point-like masses with masses in range 
$10^3 - 10^4 M_\odot$, reproduce quite nicely the observed inhomogeneous structure 
of the universe \cite{schaye}. Of course, the masses used in these simulations were not chosen 
for any fundamental reason but for computational convenience. On the other hand, 
the fact that these numerical simulations provide a good account for observed large 
scale cosmic structures  provides a kind of existence proof that at least in first approximation 
IM MACHOs can explain DM. 

\bigskip

\begin{figure}[htb]
\includegraphics{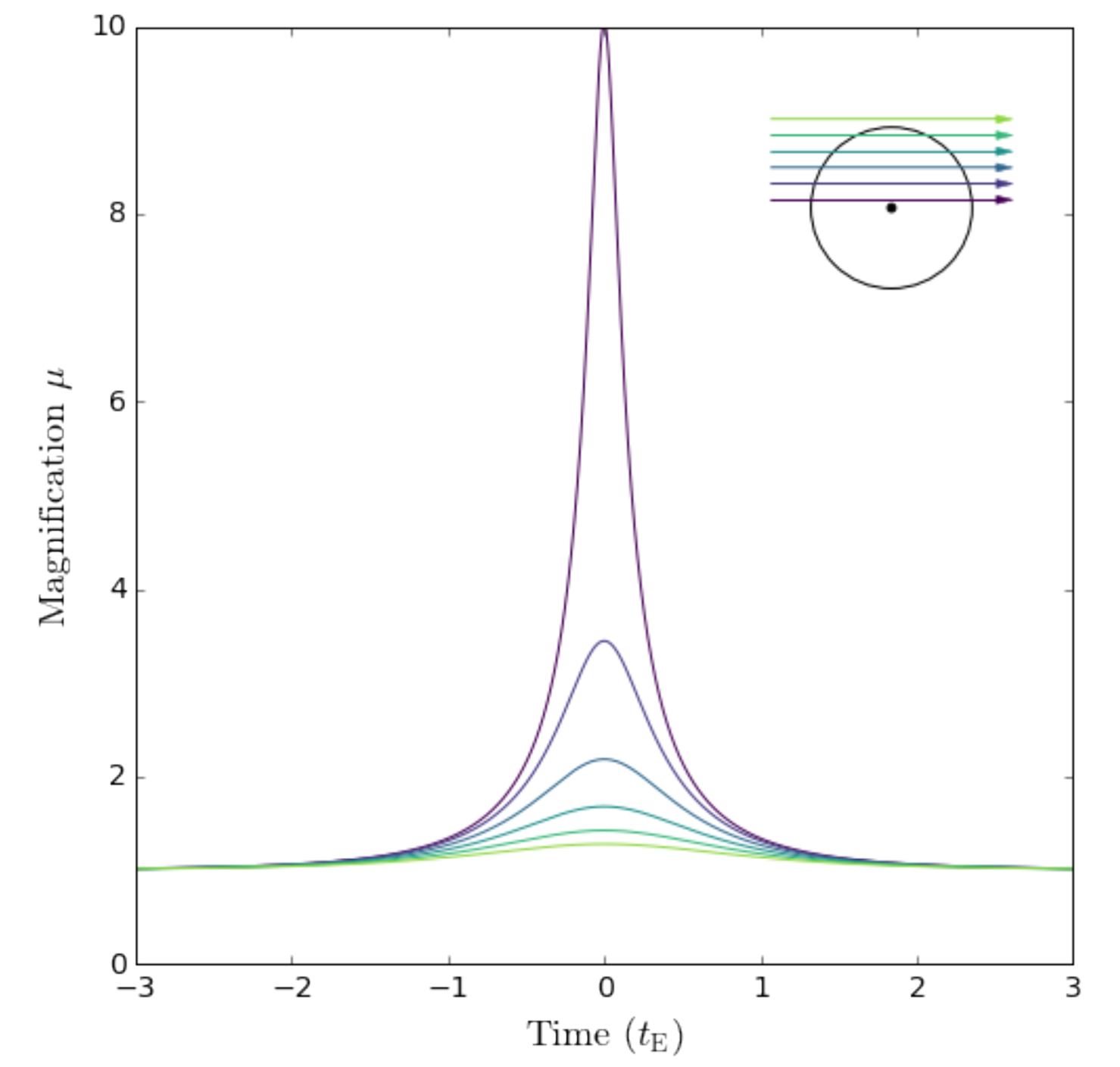}
\caption{\it Microlensing light curves as a function of time for the MACHO to 
cross the Einstein ring and the MACHO impact parameter seen by an Earth 
observer on the Earth. The circle is the Einstein radius. Based on a figure
by Penny Sackett published in Sass-Fee Advanced Course 33.}
\label{figure1}
\end{figure}

\bigskip

\noindent
It is also worth noting that when the DM gravitational dynamics simulations are 
amended by adding gas dynamics the observed morphologies of galaxies can be 
explained if the $10^3 - 10^4M_{\odot}$ point masses are supplemented with 
$10^5M_{\odot}$ point masses \cite{schaye}. This is consistent with the idea 
that the supermassive compact objects at the centers of galaxies may be primordial 
in origin. It might also be mentioned that there are theoretical arguments based 
on observed entropy of the universe \cite{chapline3,frampton6} that DM should consist of massive 
compact objects.

\section{Searching for IM MACHOs}

\bigskip

\noindent
We are interested in the gravitational lensing of stars just outside our galaxy  
(e.g. in the Magellanic Clouds) by IM MACHOs within our Galaxy. The possible 
use of gravitational microlensing to search for MACHOs with masses much higher 
than $10M_\odot$, was first discussed in detail by Gould \cite{gould2}. From the perspective 
of an Earth observer the microlensing event sought in these searches is simply 
described as a characteristic transient brightening of the source star, where the 
time dependence arises from the motion of the lens relative to the observer's line 
of sight to the source. Some brightness amplification curves for some selected 
values of the projected distance of the passing MACHO from the star being 
imaged are shown in Fig.1. 
               
\bigskip

\noindent  
The x-axis in Fig.1 is the Einstein ring crossing time  $t_E \equiv R_E/v$, 
where $R_E = (4GMD/c^2)^{1/2}$ is the radius of the Einstein ring and 
v is the MACHO velocity perpendicular to the line of sight to the star, 
which for a MACHO in the galactic halo can be approximated as the virial 
velocity for the halo; i.e. $\sim 210 km/s$. D is related to the distance to 
the star ds and distance from the Earth to the microlensing object $d_L$ (L=lens)
by $D = d_Ld_s/(d_L+d_s)$. For sources in the Magellanic Clouds D would 
typically be $\approx 10$ kpc. This tells us that the duration of the microlensing 
brightening will be on the order $t_E \approx 0.2 \sqrt{(M/M_\odot)}$ years, 
so for $100M_\odot < M <10^4M_\odot$ one would have $2yr< t_E < 20yr$. 
Since the maximum practical duration of a survey is perhaps 10 years, the 
first challenge to be faced by our proposal is that many of the events we are 
interested in may have traversed only a fraction of their full light curve 
during the survey time. 

\bigskip

\begin{figure}[htb]
\includegraphics{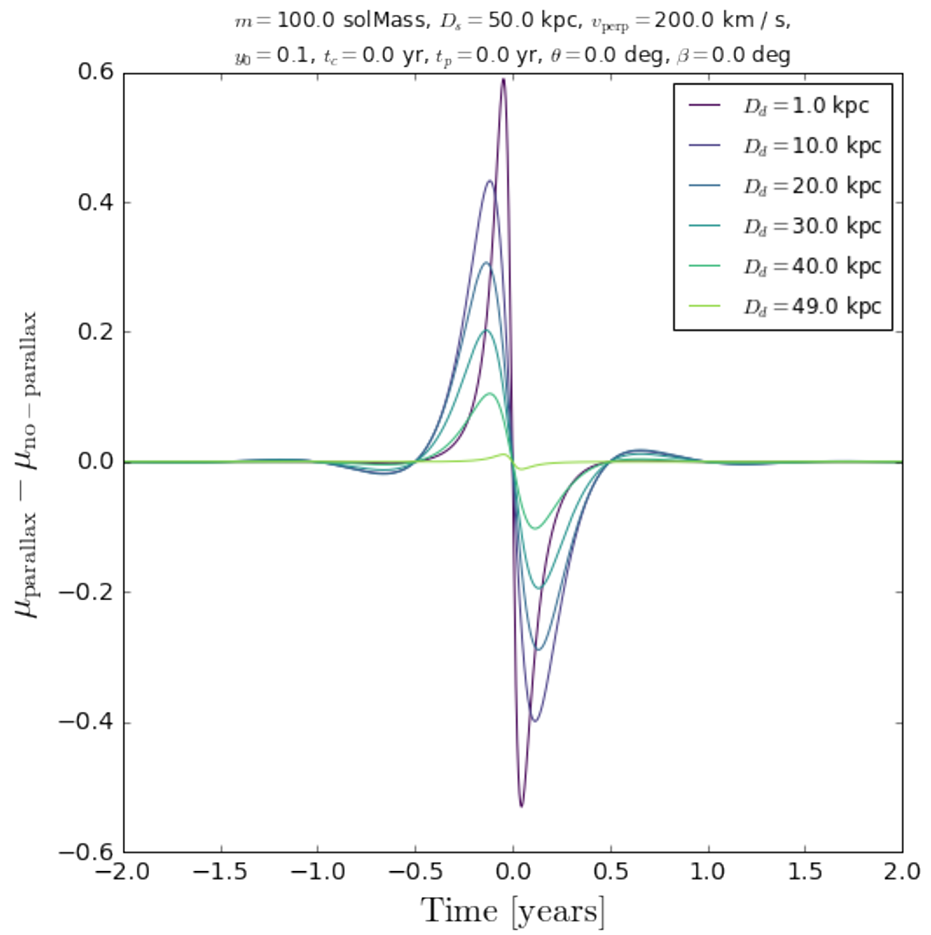}
\caption{\it An illustration of the differential effect of Earth's orbital motion around the sun 
on observed micro-lensing curves as a function of the lens distance.}
\label{figure1}
\end{figure}
         
\bigskip

\noindent
A pertinent question is how many micro-lensing events could one hope to 
detect during a survey of say $5$ years duration? The number of microlensing 
events in progress at any time, defined as those with a source inside the 
Einstein ring of the lens, is given by $N_{event} = N\tau$ where N is the 
number of LMC sources stars being monitored and $\tau$ is the micro-lensing 
optical depth. $\tau$ is independent of M, depending only on the structure of 
the Galaxy and the assumption that the dark matter which determines its 
kinematics consists of MACHOs. A value of $\tau \sim 5 \times 10^{-7}$ is a reasonable estimate 
for the optical depth to the Magellanic Clouds \cite{alcock}.  Determining N is more difficult, 
since it depends not only on the actual number of stars in the survey are to some 
limiting magnitude, but also on the capabilities of the algorithms to identify individual 
stars. As reference points, the MACHO survey \cite{alcock} monitored $10^7$ stars in the 
central region of the LMC while the OGLE survey \cite{wyrzykoeski} monitored $8 \times 10^7$ 
stars in both Magellanic Clouds.  Both surveys monitored stars brighter than visual 
magnitude 20.5, a limit which misses the bulk of the Magellanic Cloud's main sequence 
stars. With the 4 meter Blanco telescope one could probably reach $V = 23.5$ mag, 
while with the 8 meter LSST one could do even better, going to at least $V = 24.5$ mag. 
If we assume that we can resolve all the stars in the Magellanic Clouds down to $V = 24.5$ 
mag, then conservatively $N > 10^8$ could yield $N_{event} > 50$. This task will be 
challenging, given that for IM MACHOs near to the top of the mass range (1) the 
stellar brightness will vary over the survey duration by amounts which are comparable 
to the photometric errors (For the LSST: 0.001mag up to V=20mag; 0.1mag up to V=24.5mag), 
and that there will be other sources of stellar variability, both due to astrophysical effects and systematic 
errors in photometry. 

\bigskip

\noindent
For the purposes of eliminating spurious variations in stellar brightening, the previously 
used technique of looking for achromatic variation will still be useful. One advantage of 
monitoring a candidate micro-lensing event over a period longer than 1 year though is 
that this provides an opportunity to use the motion of the Earth around the Sun. As first 
suggested by Gould \cite{gould3}  it is possible to the parallax effect of the Earth's motion on 
the microlens brightening curve to not only distinguish microlensing from intrinsic 
brightening, but also say quite a bit about the phase space density of MACHOs. Fig. 2
shows the effect of the Earth's motion on the micro-lensing magnification curve as a 
function of MACHO mass. This parallax effect, first observed by the MACHO Collaboration in \cite{alcock2},
would allow one to distinguish 
microlensing brightening from intrinsic stellar variability or background sources 
(e.g. variable stars, supernova, etc.) that may exhibit brightening but no parallactic 
variability. In Figs. 2 and 3 we show how the parallax effect depends on MACHO galactic 
velocity and distance to the MACHO. Such information would be invaluable for 
confirming that the micro-lensing MACHOs have the properties expected if they are 
responsible for the DM halo of our galaxy.  Further details about the parallax effect
are provided in \cite{gould4}.

\bigskip

\noindent
It was also suggested by Gould\cite{gould5} that the masses of the microlensing objects can in 
principle also be estimated by directly observing the Einstein ring. As candidates 
are identified, their compact nature can in principle be unambiguously determined 
using adaptive optics imaging from the ground. For masses between $10^3$ and 
$10^6M_\odot$, the angular size of the Einstein ring is 
$0.025 < \theta_E < 0.8$ arcsec. This is comparable with the demonstrated 
$0.032$ arcsec angular resolution of Southern Hemisphere Magellan adaptive 
optics system, so direct imaging of the Einstein ring for the most massive 
IM MACHOs may already be a possibility. However, direct imaging of the Einstein 
ring over the full mass range (1) will probably have to wait until first light for the 
Thirty Meter or Giant Magellan Telescopes. 

\bigskip

\begin{figure}[htb]
\includegraphics{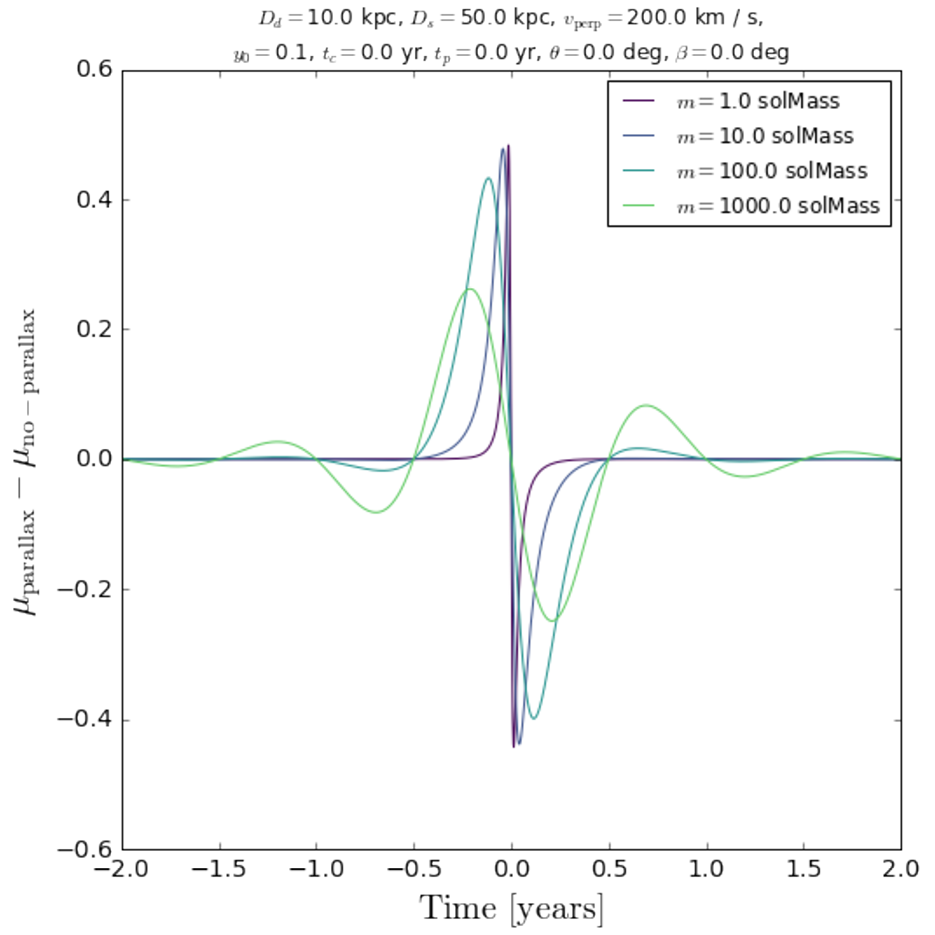}
\caption{\it An illustration of the differential effect of Earth's orbital motion around the sun 
on observed micro-lensing curves as a function of MACHO  mass. This effect allows one 
to determine the MACHO mass spectrum.}
\label{figure1}
\end{figure}
\bigskip

\noindent
Of course, in order to identify an incipient microlensing event the brightness of each individual 
star with must be recorded with sufficient accuracy with each exposure to determine if the star 
has brightened in a manner consistent with micro-lensing. This accuracy will be limited by 
the duration of each exposure and how often the exposure is made; i.e. the "cadence". Among 
existing astronomical facilities, the Blanco Dark Energy Camera in Chile may be the best 
candidate for achieving the $\sim 1\%$ level of photometry accuracy that would needed to 
identify IM MACHO candidates in the next few years. The LSST, which is expected to be able 
to measure stellar brightness with an accuracy of $0.1$ magnitude down to its limiting 
magnitude $V = 24.5$, would almost certainly be capable of identifying IM MACHO candidates 
after it becomes available as a user facility. Using the LSST to cover the Magellanic Clouds, we 
estimate that a $30$ second exposure every 2 weeks ought to suffice to produce $> 100$ 
microlensing candidates over a period of $\sim 5$ years after first light if DM does indeed 
consist of IM MACHOs.

\section{Summary}

As noted in our introduction the recent LHC results have cast doubt on the hypothesis that DM 
consists of WIMPs, and have refocused attention on the idea that primordial MACHOs may 
be responsible for DM. While there is some possibility that a DM density of MACHOs could 
be discerned in the statistics of gravitationally bound structures or an inhomogeneity of the 
cosmic microwave background, methods based on the gravitational lensing of light offer a 
more direct approach to determining whether for DM consists of massive MACHOs. Although it may be possible to detect
IMMACHOs using existing instruments such as Blanco DECam and Magellanic AO system, the 
LSST when it comes online circa 2022 will provide a definitive capability
for rapidly detecting and determining the mass spectrum
of the DM MACHOs, if they exist.

\bigskip

\section*{Acknowledgements}

We are grateful to Tim Axelrod for many helpful discussions. In particular we are grateful to 
Tim Axelrod for pointing the possibility of using the Magellan AO system to detect the Einstein 
rings of IM MACHOs.  We would also like to thank William Dawson for 
pointing out the signature characteristics of the parallax effect for
IMMACHOs and help with preparing the Figures.

\end{document}